\documentclass[onecolumn,preprint,showkeys,preprintnumbers,amsmath,amssymb,superscriptaddress,aps,prb]{revtex4-2}
\usepackage{graphicx}
\usepackage{dcolumn}
\usepackage{bm}
\usepackage{soul}
\usepackage{color}
\usepackage{epstopdf}
\usepackage[version=3]{mhchem}
\usepackage{lipsum}
\usepackage[outercaption]{sidecap}
\usepackage{floatrow}
\usepackage{hyperref}
\usepackage{appendix}
\usepackage{mathrsfs}
\usepackage{multirow}
\usepackage{booktabs}

\begin{document}

\title{Effects of magnetic field and structural parameters on multi-photon absorption spectra in Morse quantum wells with electron-phonon interactions}
\author{Tran Ky Vi}
\author{Nguyen Anh Tuan}
\affiliation{Faculty of Semiconductor Technology, Dai Nam University, Hanoi, Vietnam.}
\author{Le Nguyen Dinh Khoi}
\author{Nguyen Quang Hoc}
\affiliation{Faculty of Physics, Hanoi National University of Education, Hanoi, Vietnam.}%
\author{Anh-Tuan Tran}
\email{trananhtuan\_sdh21@hus.edu.vn}
\affiliation{Department of Theoretical Physics, Faculty of Physics, VNU University of Science, Hanoi, Vietnam.}%
\date{\today}

\begin{abstract} 
We present a systematic theoretical study of the multi-photon nonlinear optical absorption properties of a $\text{GaAs/A}{{\text{l}}_{x}}\text{G}{{\text{a}}_{1-x}}\text{As}$ based quantum well (QW) structure with Morse confinement potential  under the influence of a magnetic field. Based on the stationary states due to the electron confinement in Morse QWs and the Landau levels obtained by solving the Schrodinger equation in the effective mass approximation, we have developed calculations for the optical absorption power with MPA using second-order perturbation theory. Our model accounts for electron-phonon interactions and considers both optical and acoustic phonon mechanisms in the MPA process. Our findings show that the one-photon absorption (1PA) peaks are larger and appear to the right of the two-photon absorption (2PA) peaks, whereas 2PA peaks are larger and occur to the right of three-photon absorption (3PA) peaks. The resonance peak positions follow the magneto-phonon resonance condition and are temperature-independent. Increasing the magnetic field and aluminum concentration induces a blue shift in the absorption spectra, whereas increasing the QW width leads to a red shift. Variations in magnetic field, aluminum concentration, and QW width also affect the peak intensities and full-width at half maximum (FWHM), with increasing values of the former two enhancing the FWHM, while expanding the QW width reduces it. Thermal excitations increase peak intensity without shifting their positions. Our study highlights the significance of nonlinear absorption processes (2PA, 3PA) in understanding optical absorption, despite their smaller FWHM compared to linear absorption (1PA). Overall, the Morse QW model demonstrates promising magneto-optical properties, making it a strong candidate for future optoelectronic device applications. 
\end{abstract}

\maketitle
\section{Introduction}
Since the 1970s, the field of semiconductor physics has advanced rapidly, driven by the development of heterostructured semiconductor systems. Pioneered by Zhores I. Alferov and Herbert Kroemer \cite{nobel1,nobel2}, these systems were initially designed to enhance the speed of information transmission in photonic devices, laying the foundation for modern optoelectronics. For their groundbreaking contributions, both Alferov and Kroemer were awarded the Nobel Prize in Physics at the start of the 21st century, recognizing the pivotal role that heterostructures play in modern technology. Entering the 21st century, low-dimensional semiconductor systems continue to be widely studied due to their unique optical and electronic properties. The reason behind these interesting electronic properties is the emergence of quantum confinement effects that restrict electrons to move in certain directions. The confinement of electrons in such low-dimensional structures leads to discrete energy levels (called electric subband levels), fundamentally altering the behavior of charge carriers compared to bulk materials \cite{davies}. Among the low-dimensional semiconductor models, quantum wells (QWs) are quasi-two-dimensional semiconductor layers \cite{van3} in which the electron confinement has restriction with one-dimensional. 
Despite being first developed by L. Esaki and R. Tsu \cite{cn2} approximately 50 years ago, QWs continue to garner significant interest from physicists. This is attributed to the straightforward nature of the mathematical model and the epitaxial growth technique for superfine layers in laboratory settings \cite{cn1}. 

One of the most interesting properties in current interest in low-dimensional semiconductor systems in general and two-dimensional QW systems in particular is the influence of spatial confinement on the behavior of electrons and the control of the parameters of the confinement potential to realize new properties of the physical system \cite{cn3}. Among the QW models with different confinement potentials, the Morse QW is considered as an effective model for studying the effects of geometrical structure parameters on the magneto-optical effects of materials because of its adjustable asymmetric depth or width. Furthermore, to improve the control and manipulation of optoelectronic devices, numerous earlier studies \cite{cn4, ugan1, ugan2, ugan3, w1, m2} have looked at QW systems in electric, magnetic field, and linearly polarized electromagnetic waves (LPEMW). These studies showed that increasing external fields has significant effects on the optical properties of physics systems. For Morse confinement potential, F. Ungan \textit{et al.} \cite{m2} have clarified the significant influence of the intense laser field and the structural parameters of the Morse QW on the resonance peak position and peak intensity of the nonlinear optical rectification, total absorption, second and third harmonic generation coefficients. However, as with most current studies involving Morse confinement potential \cite{m3, m4}, the role of electron-phonon interactions has not been fully addressed. 

In the presence of a magnetic field, electron motion in the perpendicular plane is quantized into Landau levels, leading to transitions between these levels via multi-photon absorption (MPA) processes of LPEMW, accompanied by phonon absorption/emission. This phenomenon, called magneto-phonon resonance or cyclotron-phonon resonance, has been studied in detail both theoretically \cite{ltcprbulk1,ltcprbulk2,ltcpr3,ltcpr4} and experimentally \cite{tncpr1, tncpr2} in bulk semiconductor systems as well as in QWs with various confinement models. The study of magneto-phonon resonance detection conditions provides important information about the effective mass, $g$ factor, the difference between energy levels, and scattering processes \cite{epmt}. Several geometric models of confinement potentials have been used by many authors in studying the magneto-optical properties of materials such as square potential \cite{sq}, parabolic potential \cite{pqw1,pqw3}, semi-parabolic potential \cite{spqw,smpqw,afqw,pqw}, hyperbolic potential \cite{hqw,oapwt2}, triangular potential \cite{tqw}, Gaussian potential \cite{gqw}, Pöschl–Teller potential \cite{mptqw, ptqw, ptqw2}, etc. In the above studies, the authors presented calculations based on perturbation theory of the dependence of the magneto-optical absorption coefficient and the absorption spectral linewidth or full-width at half maximum (FWHM) on the geometrical structure parameters of various confinement potentials taking into account both phonon absorption and emission processes. The results indicate blue-shift behavior and increase in intensity of the peaks as the magnetic field increases. Although the results obtained are interesting, the above studies have not been performed in Morse QWs. In addition, previous studies mainly considered the interaction of electrons with optical phonons; the electron-acoustic phonon interaction mechanism, though important, is still lacking.

In this paper, we calculate the optical absorption power (OAP) in Morse QWs, taking into account both phonon interaction mechanisms within the framework of second-order perturbation theory. This approach allows us to examine in detail the contribution of MPA processes to the electron absorption spectra as well as to shed light on the electron-phonon interaction picture under the influence of external fields and Morse parameters of QWs. A notable distinction from previous studies on other confinement models is our specific consideration of the three-photon absorption (3PA) contribution, which is typically regarded as negligible compared to the contributions of one-photon absorption (1PA) (linear absorption) and two-photon absorption (2PA). Furthermore, the present work is quite different from previous works because we consider the variation of OAP and provide new analytical expressions for the dependence of FWHM on the parameters of Morse potential as well as the external fields in the electron-acoustic phonon interaction mechanism. The rest of the paper is organised as follows: In the first part of Sect. 2, we present the solution of the Schrodinger equation in Morse QWs in the effective mass approximation. In the second part of Sect. 2, we present the derivation of the analytical expression of the OAP using perturbation theory. In Sec. 3, on the basis of the obtained analytical expressions of the OAP, we perform numerical calculations, derive the absorption spectra, and calculate the FWHM by the profile method \cite{graphene}. Physical discussions and detailed analysis of the analytical results are also presented in this section. The conclusions are given in Sec. 4.

\section{Model and formalism}
\subsection{Electronic states in Morse QW under applied magnetic field}
\begin{figure}[!htb]
    \centering
    \includegraphics[width=1\linewidth]{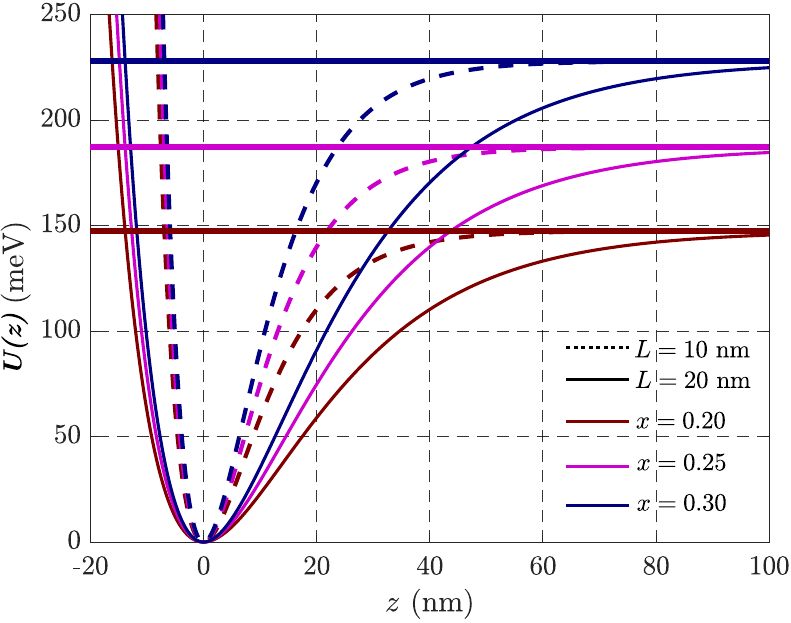}
    \caption{The potential shape are displayed for two distinct values of the well width and three different values of the Aluminium concentration.}
    \label{fig1}
\end{figure}

We consider the typical $\text{GaAs}/\text{A}{{\text{l}}_{x}}\text{G}{{\text{a}}_{1-x}}\text{As}$ QW in which the electron moves freely in the (xy) plane and is confined along the z axis by the Morse potential as follows \cite{m0,m1}
\begin{align}
    U\left( z \right)={{U}_{0}}{{\left[ 1-\text{exp}\left( -\frac{z}{L} \right) \right]}^{2}},
\end{align}
with ${{U}_{0}}=0.6\times \left( 1155\times x+370\times {{x}^{2}} \right)\left( \text{meV} \right)$ ($x$ being the Aluminium concentration) is the barrier height of the Morse QW, while $L$ is the well width. Some potential profiles of Morse QWs with different parameter values of well width and aluminum concentration are given in Fig. \ref{fig1}. From Fig. \ref{fig1}, it can be seen that as the aluminum concentration increases, the barrier height increases, leading to an increase in the number of bound states. Therefore, aluminum concentration plays an important role in influencing the electronic structure and optical absorption properties in Morse QWs. 

\begin{figure}[!htb]
    \centering
    \includegraphics[width=1\linewidth]{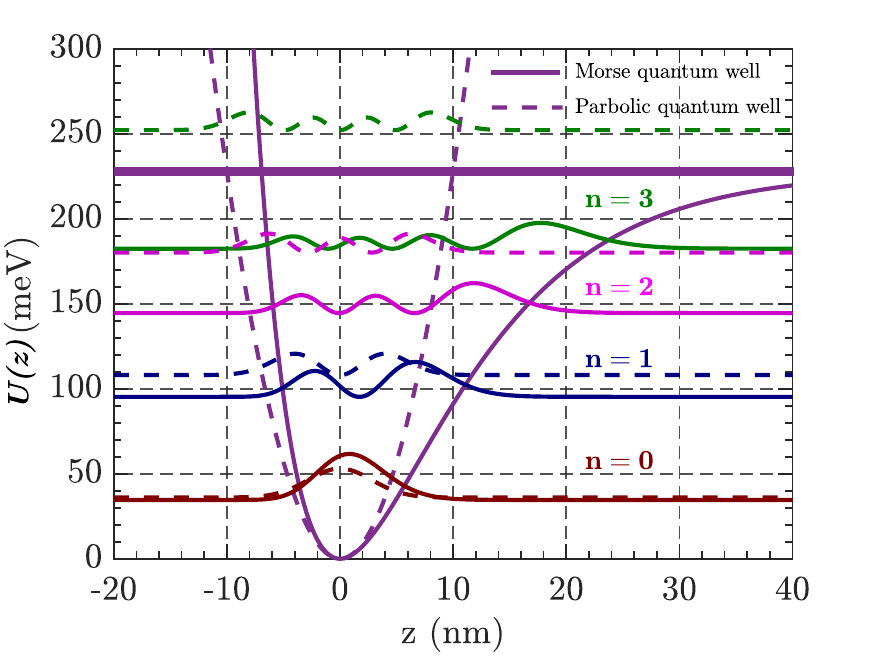}
    \caption{Confinement potential profile and the probability densities of the four bound states as a function of the growth direction for both Morse and symmetric parabolic QWs. Here, $x=0.3, L=10\text{nm}$.}
    \label{fig2}
\end{figure}

When a static magnetic field $\mathbf{B}=\left( 0,0,B \right)$ is applied perpendicularly  to the surface layers (x-y plane) with the Landau gauge $\mathbf{A}=\left( 0,Bx,0 \right)$, the Schrödinger equation for electrons in Morse QWs is as follows 
\begin{align}
    \left[ -\frac{{{\hbar }^{2}}}{2{{m}_{e}}}{{\left( \mathbf{k}+\frac{e}{\hbar }\mathbf{A} \right)}^{2}}+U\left( z \right) \right]{{\Psi }_{\text{N},\text{n},{{k}_{y}}}}\left( \mathbf{r} \right)={{\mathcal{E}}_{\text{N},\text{n}}}{{\Psi }_{\text{N},\text{n},{{k}_{y}}}}\left( \mathbf{r} \right),
\end{align}
where, $e$, and ${{m}_{e}}=0.067m_0$ are the charge and effective mass of electrons, with $m_0$ being the electron rest mass. The energy levels and wave-functions of electrons are defined as follows \cite{van1}
\begin{align}
  & {{\Psi }_{\text{N},\text{n},{{k}_{y}}}}\left( \mathbf{r} \right)=\frac{\exp \left( i{{k}_{y}}y \right)}{\sqrt{{{L}_{y}}}}{{\Phi }_{\text{N}}}\left( x-{{x}_{0}} \right){{\phi }_{\text{n}}}\left( z \right), \\ 
 & {{\Phi }_{\text{N}}}\left( x-{{x}_{0}} \right)=\frac{\exp \left[ -{{\left( x-{{x}_{0}} \right)}^{2}}/2\ell _{B}^{2} \right]}{\sqrt{{{2}^{\text{N}}}\text{N}!\sqrt{\pi }{{\ell }_{B}}}}{{\mathcal{H}}_{\text{N}}}\left( \frac{x-{{x}_{0}}}{{{\ell }_{B}}} \right), \\ 
 & {{\mathcal{E}}_{\text{N},\text{n}}}=\left( \text{N}+\frac{1}{2} \right)\hbar {{\omega }_{B}}+{{\mathcal{E}}_{\text{n}}}, 
\end{align}
here, $\mathbf{k}=\left( {{k}_{x}},{{k}_{y}},{{k}_{z}} \right)$ is the electron wave vector. ${{\ell }_{B}}=\sqrt{{\hbar }/{eB}\;}$ is the magnetic length, $\text{N}=0,1,2,\ldots $ presents the Landau index, and ${{x}_{0}}=-{{{k}_{y}}}/{{{m}_{e}}{{\omega }_{B}}}\;$ is the coordinate of the center of the normalized harmonic oscillation wave function ${{\Phi }_{\text{N}}}\left( x-{{x}_{0}} \right)$, with ${{\omega }_{B}}={eB}/{{{m}_{e}}}\;$ being the cyclotron frequency. ${\mathcal{H}}_{\mathrm{N}}\left(x\right)$ is the N-th Hermite polynomials. ${{\phi }_{\text{n}}}\left( z \right)$, and ${{\mathcal{E}}_{\text{n}}}$ are respectively the wave function and subband energy due to confinement of Morse QW obtained from solving Schrodinger equation \cite{m0, m1}
\begin{align}
    &\frac{{{d}^{2}}{{\phi }_{\text{n}}}\left( z \right)}{d{{z}^{2}}}+\frac{2{{m}_{e}}}{{{\hbar }^{2}}}\left[ {{\mathcal{E}}_{\text{n}}}-U\left( z \right) \right]{{\phi }_{\text{n}}}\left( z \right)=0, \\ 
    &{{\phi }_{\text{n}}}\left( z  \right)={{\mathscr{C}}_{\text{n}}}{{\xi }^{\text{j}}}\text{exp}\left( -\frac{\xi }{2} \right)\mathcal{L}_{\text{n}}^{2\text{j}}\left( \xi  \right), \\
    &{{\mathcal{E}}_{\text{n}}}=\left[ \left( \text{n}+\frac{1}{2} \right)-\frac{1}{2\lambda }{{\left( \text{n}+\frac{1}{2} \right)}^{2}} \right]\hbar {{\omega }_{z}},
\end{align}
in which, $\xi =2\lambda \exp \left( {-z}/{L}\; \right)$, $\lambda ={L\sqrt{2{{m}_{e}}{{U}_{0}}}}/{\hbar }\;$, and $\text{j}=\lambda -\text{n}-{1}/{2}\;$ with $\text{n}=0,1,2,...,\left[ \lambda -{1}/{2}\; \right]$ being the electric subband. $\left[ x \right]$ is the greatest integer function, which means this function returns the nearest integer value close to $x$. ${{\mathscr{C}}_{\text{n}}}=\sqrt{{2\text{j}\Gamma \left( \text{n}+1 \right)}/{\Gamma \left( \text{2}\lambda -\text{n} \right)}\;}$ is the normalization coefficient. ${{\omega }_{z}}={\hbar \sqrt{{2{{U}_{0}}}/{{{m}_{e}}}\;}}/{L}\;$ is the confinement frequency of Morse QWs. $\Gamma\left( x \right)$ and $\mathcal{L}_{\text{n}}^{2{\text{j}}}\left( x \right)$ are the Gamma functions and associated Laguerre polynomials of degree ${\text{n}}$ in ${2{\text{j}}}$, respectively. In Fig. \ref{fig2}, we present the results of numerical calculations subband energies $\mathcal{E}_{\mathrm{n}}$ and probability densities ${{\left| {{\phi }_{\text{n}}}\left( z \right) \right|}^{2}}$ corresponding to different potential profile cases for both Morse and symmetric parabolic QWs. Furthermore, it can be seen from Fig. \ref{fig2} that unlike the parabolic QW or other QW models with infinite barriers \cite{pqw, ptqw,gqw,hqw,tqw,spqw}, the bound states in the Morse QW are finite, namely the maximum number of bound states is ${{\text{n}}_{\text{max}}}=\left[ \lambda -{1}/{2}\; \right]$. 

\subsection{Optical absorption power in Morse QWs} 

We examine the electron-phonon system in Morse QWs subjected to the influence of the time-dependent fields of a linearly polarized electromagnetic wave with an angular frequency $\Omega$ (on the order of terahertz) and an intensity ${\mathscr{E}}_0$, propagating along the x axis. To evaluate the optical absorption properties of electrons in Morse QWs, we use the well-known expression for the OAP as follows \cite{pqw1, pqw2}
\begin{align}\label{p}
{\mathscr{P}}=\frac{{\mathscr{E}}_{0}^{2}\sqrt{{{\kappa }_{0}}}}{8\pi }\sum\limits_{{{\zeta }_{1}},{{\zeta }_{3}}}{\mathscr{W}_{{{\zeta }_{1}},{{\zeta }_{3}}}^{\varsigma  ,\nu ,\ell }{{\mathscr{F}}_{{{\zeta }_{1}}}}\left( 1-{{\mathscr{F}}_{{{\zeta }_{3}}}} \right)},
\end{align}
here, ${{\mathscr{F}}_{{{\zeta }_{1}}}}$ is the equilibrium distribution function for electrons in the presence of a magnetic field \cite{pqw2}, and $\kappa_0$ is is the dielectric constant of the material. Eq. \eqref{p} is derived from the common definition of the optical absorption power $\mathscr{P}=\frac{{\mathscr{E}}_{0}^{2}}{2}\operatorname{Re}\left[ \sigma \left( \Omega  \right) \right]$, with $\operatorname{Re}\left[ \sigma \left( \Omega  \right) \right]={nc\alpha \left( \Omega  \right)}/{4\pi }\;$ is the real part of the optical conductivity, in which $\alpha \left( \Omega  \right)$ is the absorption coefficient, $n$ is the refractive index, and $c$ is the velocity of light. Finally, from the analytical expression of the absorption coefficient in \cite{ltcpr3}, we obtain the optical absorption power in Eq. \eqref{p}. One thing to note here is that the electron gas system we consider is assumed to be non-degenerate, so we can ignore ${{\mathscr{F}}_{{{\zeta }_{3}}}}$ when calculating the OAP \cite{seeger}. This leads to Eq. \eqref{p} which has the same form as in previous studies \cite{pqw1, pqw2}.   

For the multi-photon absorption process, since it involves interactions between three particle, namely electron, photon and phonon, we use the second-order perturbation theory as in the previous works of W. Xu \textit{et al.} \cite{w1} and K. Seeger \cite{seeger}, in which an intermediate or virtual state $\left| {{\zeta }_{2}} \right\rangle$ is assumed to exist between the initial state $\left| {{\zeta }_{1}} \right\rangle$ and the final state $\left| {{\zeta }_{3}} \right\rangle$. Using the second-order Fermi golden rule, one can obtain the transition probability for the electron-phonon-photon scattering mechanism from the initial state $\left| {{\zeta }_{1}} \right\rangle \equiv \left| \text{N},\text{n} \right\rangle $ to the final state $\left| {{\zeta }_{3}} \right\rangle \equiv \left| {{\text{N}}^{\prime }},{{\text{n}}^{\prime }} \right\rangle$, taking into account the MPA process as follows \cite{w1, w2}
\begin{align}\label{xs}
    \mathscr{W}_{{{\zeta }_{1}},{{\zeta }_{3}}}^{\varsigma ,\nu ,\ell }=\frac{2\pi }{\hbar }\sum\limits_{{{\zeta }_{2}},\mathbf{q}}{{{{{{\left| {{\mathcal{M}}_{{{\zeta }_{1}}\to {{\zeta }_{2}}\to {{\zeta }_{3}}}}\left( \mathbf{q} \right)\right|}}^{2}}}}\sum\limits_{\ell }{\frac{{{\left( {{a_{0}}}q \right)}^{2\ell }}}{{{2}^{2\ell }}{{\left( \ell ! \right)}^{2}}}\delta \left( {{\mathcal{E}}_{{{\zeta }_{3}}}}-{{\mathcal{E}}_{{{\zeta }_{1}}}}+\varsigma \hbar \omega _{\mathbf{q}}^{\nu }-\ell \hbar \Omega  \right),}}
\end{align}
here, the summation is over all possible virtual or intermediate states $\left| {{\zeta }_{2}} \right\rangle$. Virtual transitions are usually non-energy-conserving. First, $\left| {{\zeta }_{1}} \right\rangle$  makes a non-energy-conserving transition to virtual state $\left| {{\zeta }_{2}} \right\rangle$ (photon absorption), then $\left| {{\zeta }_{2}} \right\rangle$ makes a non-energy-conserving transition to $\left| {{\zeta }_{3}} \right\rangle$ (phonon absorption or emission), between $\left| {{\zeta }_{3}} \right\rangle$ and $\left| {{\zeta }_{1}} \right\rangle$ there is overall energy conservation \cite{sakurai, nazarov}. The physical meanings of $\mathscr{W}_{{{\zeta }_{1}},{{\zeta }_{3}}}^{\varsigma ,\nu ,\ell }$ in Eq. \eqref{xs} are as follows: $\varsigma =+1$ or $-1$ corresponds to the phonon emission or absorption process. $\mathbf{q}=\left( {{q}_{\bot }},{{q}_{z}} \right)$, and $\hbar \omega _{\mathbf{q}}^{\nu }$ are phonon wave vector, and phonon energy with different scattering mechanisms ($\nu=\text{op}$ or ac corresponds to the optical or acoustic phonon scattering). $a_0 =7.5$nm denotes the laser-dressing parameter \cite{ptqw, mptqw, smpqw}. The Dirac delta function $\delta \left( {{\mathcal{E}}_{{{\zeta }_{3}}}}-{{\mathcal{E}}_{{{\zeta }_{1}}}}+\varsigma \hbar \omega _{\mathbf{q}}^{\nu }-\ell \hbar \Omega  \right)$ defines energy conservation in the transition from the initial state to final state, accounting for the $\ell$-photon absorption process. ${{\mathcal{M}}_{{{\zeta }_{1}}\to {{\zeta }_{2}}\to {{\zeta }_{3}}}}\left( \mathbf{q} \right)$ is the matrix element representing the transition of an electron from the initial state to the final state  through the intermediate state $\left| {{\zeta }_{2}} \right\rangle \equiv \left| {{\text{N}}^{\prime \prime }},{{\text{n}}^{\prime \prime }} \right\rangle $  
\begin{align}\label{ytmt2}
    {{\left| {{\mathcal{M}}_{{{\zeta }_{1}}\to {{\zeta }_{2}}\to {{\zeta }_{3}}}}\left( \mathbf{q} \right) \right|}^{2}}={{\left| \frac{\mathcal{M}_{{{\zeta }_{1}}\to {{\zeta }_{2}}}^{\left( \text{rad} \right)}}{\hbar \Omega } \right|}^{2}}{{\left| \mathcal{M}_{{{\zeta }_{2}}\to {{\zeta }_{3}}}^{\varsigma, \nu }\left( \mathbf{q} \right) \right|}^{2}},
\end{align}
in which, the electron-photon interaction matrix element can be written as \cite{slchuan} 
\begin{align}\label{ytmt}
    {{\left| \mathcal{M}_{{{\zeta }_{1}}\to {{\zeta }_{2}}}^{\left( \text{rad} \right)} \right|}^{2}}=\frac{{{\Omega }^{2}}{\mathscr{A}}_{0}^{2}}{4}{{\left| \mathbf{n}\cdot e\left\langle  {{\zeta }_{2}} \right|\mathbf{r}\left| {{\zeta }_{1}} \right\rangle  \right|}^{2}},
\end{align}
here, $\mathbf{n}$ is the polarized vector, and ${{{\mathscr{A}}}_{0}}={{{\mathscr{E}}_{0}}}/{\Omega }\;$ is the amplitude of the vector potential of the LPEMW. $\left\langle  {{\zeta }_{2}} \right|\mathbf{r}\left| {{\zeta }_{1}} \right\rangle $ is the dipole moment matrix element, calculated assuming the LPEMW propagates along the x-axis as follows \cite{ptqw, mptqw}
\begin{align}
   {{\mathcal{Q}}_{{{\text{N}}^{\prime \prime }},\text{N}}}\equiv {{\left| \langle {{\zeta }_{2}}|x\left| {{\zeta }_{1}} \right\rangle  \right|}^{2}}={{\left[ \frac{{{\ell }_{B}}}{\sqrt{2}}\left( \sqrt{\text{N}}{{\delta }_{{{\text{N}}^{\prime \prime }},\text{N}-1}}+\sqrt{\text{N}+1}{{\delta }_{{{\text{N}}^{\prime \prime }},\text{N}+1}} \right)+{{x}_{0}}{{\delta }_{{{\text{N}}^{\prime \prime }},\text{N}}} \right]}^{2}}.
\end{align}

Besides, in Eq. \eqref{ytmt2}, the
part of the electron-phonon interaction is given as follows \cite{epmt}
\begin{align}\label{mep}
   {{\left| \mathcal{M}_{{{\zeta }_{2}}\to {{\zeta }_{3}}}^{\varsigma ,\nu }\left( \mathbf{q} \right) \right|}^{2}}={{\left| \mathcal{C}_{\mathbf{q}}^{\nu } \right|}^{2}}{{\left| {{\mathcal{I}}_{{{\text{n}}^{\prime }},{{\text{n}}^{\prime \prime }}}}\left( \pm {{q}_{z}} \right) \right|}^{2}}{{\left| {{\mathcal{J}}_{{{\text{N}}^{\prime }}\text{,}{{\text{N}}^{\prime \prime }}}}\left( {{\mathbf{q}}_{\bot }} \right) \right|}^{2}}\mathcal{N}_{\mathbf{q}}^{\varsigma ,\nu },
\end{align}
where, $\mathcal{C}_{\mathbf{q}}^{\nu }$ is the electron-phonon interaction coefficient, which depends on the interaction mechanism (recall, $\nu=\text{op}$ or ac). ${{\mathcal{I}}_{{{\text{n}}^{\prime }},{{\text{n}}^{\prime \prime }}}}\left( \pm {{q}_{z}} \right)=\int\limits_{-\infty }^{+\infty }{\phi _{{{\text{n}}^{\prime }}}^{*}\left( \xi  \right)\text{exp}\left( \pm i{{q}_{z}}z \right){{\phi }_{{{\text{n}}^{\prime \prime }}}}\left( \xi  \right)dz}$ is the form factor for electron-phonon interaction, which characterizes the confinement effect of the Morse QWs along the growth direction. ${{\left| {{\mathcal{J}}_{{{\text{N}}^{\prime }},{{\text{N}}^{\prime \prime }}}}\left( {{\mathbf{q}}_{\bot }} \right) \right|}^{2}}=\frac{{{\text{N}}^{\prime \prime }}!}{{{\text{N}}^{\prime }}!}\exp \left( -u \right){{u}^{{{\text{N}}^{\prime }}-{{\text{N}}^{\prime \prime }}}}{{\left[ \mathcal{L}_{{{\text{N}}^{\prime \prime }}}^{{{\text{N}}^{\prime }}-{{\text{N}}^{\prime \prime }}}\left( u \right) \right]}^{2}}$ is induced by the electron interaction with the magnetic field, with $u={\ell _{B}^{2}q_{\bot }^{2}}/{2}\;$. $\mathcal{N}_{\mathbf{q}}^{\varsigma ,\nu }=\mathcal{N}_{\mathbf{q}}^{\nu }+\frac{1}{2}+\frac{\varsigma }{2}$, with $\mathcal{N}_{\mathbf{q}}^{\nu }$, with $\mathcal{N}_{\mathbf{q}}^{\nu }={{\left[ \exp \left( {\hbar \omega _{\mathbf{q}}^{\nu }}/{{{k}_{B}}T}\; \right)-1 \right]}^{-1}}$ is the Bose-Einstein distribution function of phonons. 

\subsubsection{Optical phonon interaction}
The deformation potential coupling coefficient for electron-optical phonon interaction ($\nu =\text{op}$) is given by \cite{pqw3,hqw,ptqw2}
 \begin{align}\label{cop}
    {{\left| \mathcal{C}_{\mathbf{q}}^{\text{op}} \right|}^{2}}=\frac{4\pi {{e}^{2}}\hbar \omega _{\mathbf{q}}^{\text{op}}}{{{V}_{0}}{{\varepsilon }_{0}}{{\mathbf{q}}^{2}}}\left( \frac{1}{{{\kappa }_{\infty }}}-\frac{1}{{{\kappa }_{0}}} \right),
 \end{align}
here, ${{\kappa }_{\infty }}=10.89-2.73\times x$, and ${{\kappa }_{0}}=13.18-3.12\times x$ \cite{para, para1} are the high- and static dielectric constant, respectively. ${{\varepsilon }_{0}}$ being the permittivity in free space. ${{V}_{0}}={{L}_{x}}{{L}_{y}}L$ is the volume of the system. $\hbar \omega _{\mathbf{q}}^{\text{op}}=\left( 36.25+1.83\times x+17.12\times {{x}^{2}}-5.11\times {{x}^{3}} \right)\text{meV}$ is the effective optical phonon energy in the barrier $\text{A}{{\text{l}}_{x}}\text{G}{{\text{a}}_{1-x}}\text{As}$ \cite{para, para1}, assuming non-dispersive optical phonons. 

For simplicity, in this study we limit ourselves to the 1PA, 2PA and 3PA processes, corresponding to $\ell = 1,2, 3$. Inserting Eqs. \eqref{xs}, and \eqref{cop} into Eq. \eqref{p}, and then doing some mathematical manipulation with the integral transformations similar to previous studies \cite{pqw1, pqw2, ptqw}, we obtain the following expression for OAP in the case of electron-optical phonon interaction 
\begin{align}\label{apop}
  & {{{\mathscr{P}}}^{\left( \text{op} \right)}}={\mathscr{P}}_{0}^{\left( \text{op} \right)}\sum\limits_{{{\text{N}}^{\prime }},{{\text{n}}^{\prime }}}{\sum\limits_{{{\text{N}}^{\prime \prime }},{{\text{n}}^{\prime \prime }}}{\sum\limits_{\text{N},\text{n}}{{{\Lambda }_{\text{N},\text{n}}}{{\mathcal{G}}_{{{\text{n}}^{\prime }},{{\text{n}}^{\prime \prime }}}}{{\mathcal{Q}}_{{{\text{N}}^{\prime \prime }},\text{N}}}\left[ {\mathscr{H}}_{\ell =1}^{\left( \text{op} \right)}+\frac{a_{0}^{2}}{8 \ell_{B}^2}{\mathscr{H}}_{\ell =2}^{\left( \text{op} \right)}+\frac{a_{0}^{4}}{144 \ell_{B}^6}{\mathscr{H}}_{\ell =3}^{\left( \text{op} \right)} \right]}}}, \\ 
 & {\mathscr{P}}_{0}^{\left( \text{op} \right)}=\frac{{{\mathscr{E}_0^2}\sqrt{{{\kappa }_{0}}}\eta {{e}^{4}}\hbar {{\omega }_{0}}a_{0}^{2}{{S}^{2}}}}{128{{\pi }^{4}}\ell _{B}^{6}{{\varepsilon }_{0}}L{{\hbar }^{2}}\Omega }\left( \frac{1}{{{\kappa }_{\infty }}}-\frac{1}{{{\kappa }_{0}}} \right), 
\end{align}
here, ${{\Lambda }_{\text{N},\text{n}}}={\exp \left( -\frac{{{\mathcal{E}}_{\text{N},\text{n}}}}{{{k}_{B}}T} \right)}/{\sum\limits_{\text{N},\text{n}}{\exp \left( -\frac{{{\mathcal{E}}_{\text{N},\text{n}}}}{{{k}_{B}}T} \right)}\ }\;$, ${{\mathcal{G}}_{{{\text{n}}^{\prime }},{{\text{n}}^{\prime \prime }}}}=\int\limits_{-\infty }^{+\infty }{{{\left| {{\mathcal{I}}_{{{\text{n}}^{\prime }},{{\text{n}}^{\prime \prime }}}}\left( \pm {{q}_{z}} \right) \right|}^{2}}d{{q}_{z}}}$ is the squared overlap integral describing crystal momentum conservation \cite{tpovl}, calculated numerically in Sec. 3.  ${\mathscr{H}}_{\ell}^{\left( \text{op} \right)}$ represents the contribution of the $\ell$-photon absorption process, given by
\begin{align}\label{hop}
    {\mathscr{H}}_{\ell}^{\left( \text{op} \right)}={\mathscr{I}}_{{{\text{N}}^{\prime }}\text{,}{{\text{N}}^{\prime \prime }}}^{\left( \ell-1 \right)}\left[\left( \mathcal{N}_{\mathbf{q}}^{\text{op}}+1 \right)\delta \left( \Delta _{\text{N},\text{n}}^{{{\text{N}}^{\prime }},{{\text{n}}^{\prime }}}+\hbar \omega _{\mathbf{q}}^{\text{op}}-\ell\hbar \Omega  \right)+{{\mathcal{N}}^{\text{op}}_{\mathbf{q}}}\delta \left( \Delta _{\text{N},\text{n}}^{{{\text{N}}^{\prime }},{{\text{n}}^{\prime }}}-\hbar \omega _{\mathbf{q}}^{\text{op}}-\ell\hbar \Omega  \right)\right],
\end{align}
in which, $\Delta _{\text{N},\text{n}}^{{{\text{N}}^{\prime }},{{\text{n}}^{\prime }}}={{\mathcal{E}}_{{{\text{N}}^{\prime }},{{\text{n}}^{\prime }}}}-{{\mathcal{E}}_{\text{N},\text{n}}}=\left( {{\text{N}}^{\prime }}-\text{N} \right)\hbar {{\omega }_{B}}+\left( {{\text{n}}^{\prime }}-\text{n} \right)\left( 1-\frac{{{\text{n}}^{\prime }}+\text{n}+1}{2\lambda } \right)\hbar {{\omega }_{z}}$ being the energy separation between the two states $\left| {{\text{N}}^{\prime }},{{\text{n}}^{\prime }} \right\rangle $ and $\left| {{\text{N}}},{{\text{n}}} \right\rangle $, and ${\mathscr{I}}_{{{\text{N}}^{\prime }}\text{,}{{\text{N}}^{\prime \prime }}}^{\left( \ell-1 \right)} = \int\limits_{0}^{+\infty }{{{u}^{\ell -1}}{{\left| {{\mathcal{J}}_{{{\text{N}}^{\prime }}\text{,}{{\text{N}}^{\prime \prime }}}}\left( u \right) \right|}^{2}}du}$, whose analytical expressions for 1PA, 2PA, and 3PA processes are calculated exactly in the Appendix A. In Eq. \eqref{hop}, term ${\left( \mathcal{N}_{\mathbf{q}}^{\text{op}}+1 \right)}/{\mathcal{N}_{\mathbf{q}}^{\text{op}}}\;$ appears due to the phonon emission/absorption condition. We will clarify the contributions of optical phonon absorption and emission processes in the numerical calculations of the absorption spectra in the next section.
\subsubsection{Acoustic phonon interaction}
The electron-acoustic phonon matrix element 
($\nu =\text{ac}$) is given by \cite{epmt}
\begin{align}\label{ac}
    {{\left| \mathcal{C}_{\mathbf{q}}^{\text{ac}} \right|}^{2}}=\frac{\hbar {{\vartheta }^{2}}q}{2\rho {{\upsilon }_{\text{S}}}{{V}_{0}}},
\end{align}
here, $\vartheta $ is the deformation potential constant, ${{\upsilon }_{\text{S}}}$ denotes the velocity of sound in the semiconductor, and $\rho $ is the density of the material. 

Using Eq. \eqref{ac}, and a straightforward calculation of integral over $\mathbf{k}_y$ and $\mathbf{q}$, we obtain the following expression for OAP in the case of electron-acoustic phonon interaction 
\begin{align}\label{apac}
  {{\mathscr{P}}^{\left( \text{ac} \right)}}&=\mathscr{P}_{0}^{\left( \text{ac} \right)}\sum\limits_{{{\text{N}}^{\prime }},{{\text{n}}^{\prime }}}{\sum\limits_{{{\text{N}}^{\prime \prime }},{{\text{n}}^{\prime \prime }}}{\sum\limits_{\text{N},\text{n}}{{{\Lambda }_{\text{N},\text{n}}}{{\mathcal{G}}_{{{\text{n}}^{\prime }},{{\text{n}}^{\prime \prime }}}}{{\mathcal{Q}}_{{{\text{N}}^{\prime \prime }},\text{N}}}\left[ \mathscr{H}_{\ell =1}^{\left( \text{ac} \right)}+\frac{a_{0}^{2}}{8 \ell_{B}^2}\mathscr{H}_{\ell =2}^{\left( \text{ac} \right)}+\frac{a_{0}^{4}}{144 \ell_{B}^4}\mathscr{H}_{\ell =3}^{\left( \text{ac} \right)} \right]}}},\\
    \mathscr{P}_{0}^{\left( \text{ac} \right)}&=\frac{{{\mathscr{E}_0^2}\sqrt{{{\kappa }_{0}}}{{e}^{2}}\eta {{\vartheta }^{2}}{{k}_{B}}T{{S}^{2}}a_{0}^{2}}}{512\pi {{\hbar }^{2}}L\Omega \rho \upsilon _{\text{S}}^{2}{{\pi }^{4}}\ell _{B}^{6}},\\ 
\mathscr{H}_{\ell }^{\left( \text{ac} \right)}&=\mathscr{I}_{{{\text{N}}^{\prime }}\text{,}{{\text{N}}^{\prime \prime }}}^{\left( \ell  \right)}\delta \left( {{\mathcal{E}}_{{{\text{N}}^{\prime }},{{\text{n}}^{\prime }}}}-{{\mathcal{E}}_{\text{N},\text{n}}}-\ell \hbar \Omega  \right), \label{hac}
\end{align}
where, we have introduced the assumption that the phonon energy is linear with the wave vector $\hbar \omega _{\mathbf{q}}^{\text{ac}}=\hbar {{\upsilon }_{\text{S}}}q$ and very small compared to the cyclotron energy, i.e. $\hbar \omega _{\mathbf{q}}^{\text{ac}}\ll \hbar {{\omega }_{B}}$. Therefore, the acoustic phonon energy can be neglected in the argument of the Dirac Delta functions. In addition, we also use approximations $\mathcal{N}_{\mathbf{q}}^{\varsigma \text{,ac}}\approx {{{k}_{B}}T}/{\hbar \omega _{\mathbf{q}}^{\text{ac}}}\;={{{k}_{B}}T}/{\hbar {{\upsilon }_{\text{S}}}q}\;$ \cite{van3,van2}. Consequently, in contrast to optical phonon interactions, the contributions of phonon absorption and emission in the expression of OAP are indistinguishable. 

The analytical expressions obtained for both the optical and acoustic phonon interactions in Eqs. \eqref{apop}, \eqref{apac} show the complex dependence of the OAP on the material parameters as well as the external fields and temperature of the system. The contributions of MPA processes have also been specifically considered. Qualitatively, from the coefficients of the $\mathscr{H}_{\ell }^{\left( \nu =\text{op/ac} \right)}$ terms, we can see that the contributions of MPA processes are smaller than those of 1PA process (linear absorption). In the next section, we will present numerical calculations to illustrate the obtained analytical results and provide physical discussions for analyzing the absorption spectra of electrons in Morse QWs with the help of computer programs. It was suggested by C. M. Van Vliet that the Dirac Delta functions in Eqs. \eqref{hop} and \eqref{hac} should be changed into Lorentzian functions because they diverge when the argument is zero. The Lorentzian spectral broadening is given by ${{\left( \Gamma _{{{\zeta }_{2}}\to {{\zeta }_{3}}}^{\varsigma,\nu } \right)}^{2}}=\sum\limits_{\mathbf{q}} {{{\left| \mathcal{M}_{{{\zeta }_{2}}\to {{\zeta }_{3}}}^{\varsigma,\nu }\left( \mathbf{q} \right) \right|}^{2}}}$ using a collision-broadening model \cite{van1, van2}.

\section{Results and discussion}
In the previous section, we obtained the analytical expressions of the OAP under the influence of a magnetic field, taking account of electron-phonon interactions. The above results will be applied to investigate numerically the OAP in Morse QWs, from which we will provide specific insights into the dependence of the absorption spectra on material parameters and external fields. The parameters used in this computational model are given as follows \cite{ptqw, hqw, gqw, pqw, pqw1, para, para1} $\vartheta =12.42\text{eV},\rho =5.31\text{g}\cdot \text{c}{{\text{m}}^{-3}},{{\upsilon }_{\text{S}}}=5.22\times {{10}^{5}}\text{cm}\cdot {{\text{s}}^{-1}}, \eta =3\times {{10}^{16}}\text{c}{{\text{m}}^{-3}}, {\mathscr{E}_0} = 4.5 \times 10^5 \text{V/m}$. In the following, we will consider only the transition between the two lowest states, i.e. $\text{N}=\text{n}=0;{{\text{N}}^{\prime }}={{\text{n}}^{\prime }}=1$ (for the extreme quantum limit) \cite{van3}, and assume that the contributions from other transitions are negligible. The numerical results based on this assumption are suitable for comparison with previous results.  

\begin{figure}[!htb]
    \centering
    \includegraphics[width=1.0\linewidth]{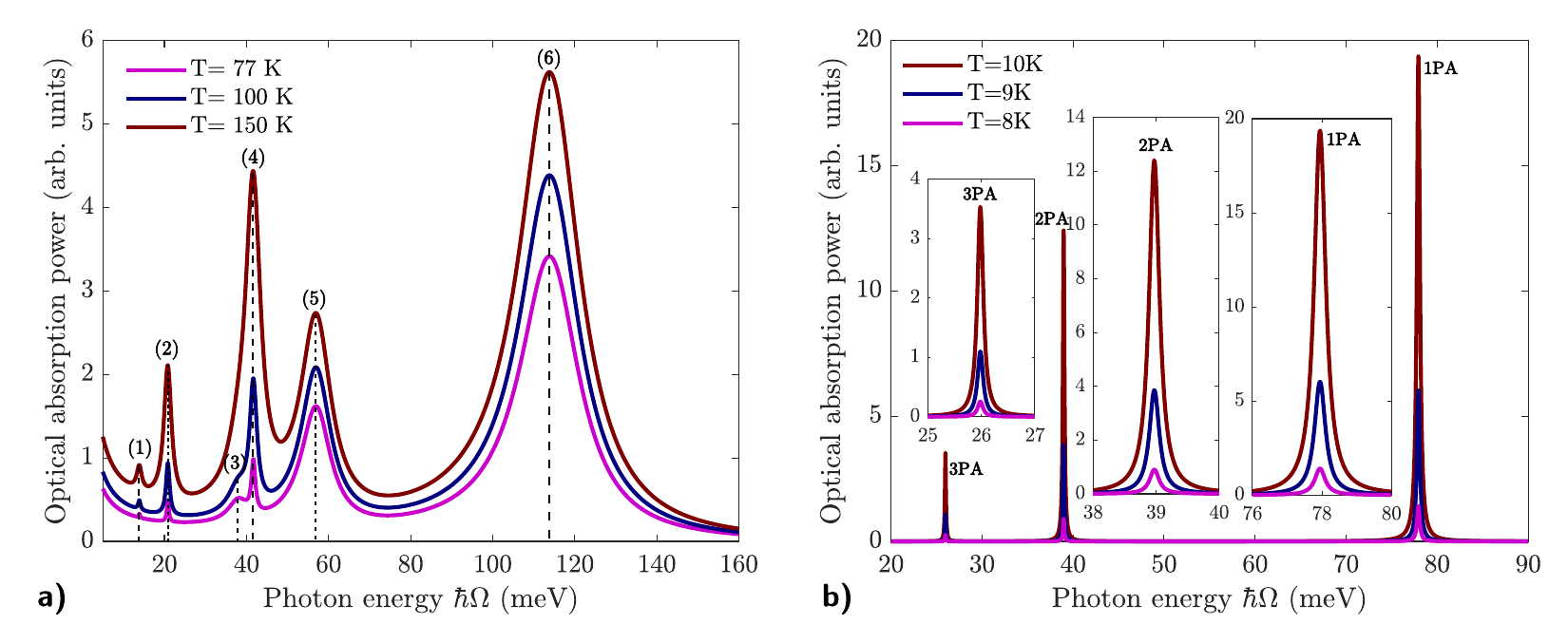}
    \caption{The variation of the OAP with the photon energy of LPEMW at three various values of the temperature with the transition between the ground state ($\text{N}=\text{n}=0$) and the first exited state (${{\text{N}}^{\prime }}={{\text{n}}^{\prime }}=1$). Here, the results are presented at $B=10$T, $L=10 \text{nm}$, and $x=0.3$ for electron-optical phonon interaction (a), and electron-acoustic phonon interaction (b).}
    \label{fig3}
\end{figure}

In Fig. \ref{fig3}, we show the OAP as a function of $\hbar\Omega$  with two interaction mechanisms at different values of the temperature. From
Figs. \ref{fig3}(a) and \ref{fig3}(b), we can see the temperature does not have an effect on the peak position, but it does effect the peak intensity. There are a total of 6 resonance peaks when the temperature is not too high, including three resonance peaks of phonon absorption and three resonance peaks of phonon emission. In addition, corresponding to each $\ell$-photon absorption process, there are three peaks, the 1PA peak is located at the position where the photon energy is the largest, then the 2PA, and 3PA peaks are located on the left and in the region of smaller photon energy. The peaks all satisfy the magneto-phonon resonance conditions (MPRC) as in previous studies of parabolic \cite{pqw, pqw1}, hyperbolic \cite{hqw}, Gaussian \cite{gqw} and Pöschl–Teller \cite{ptqw, mptqw} QWs and graphene \cite{tp2}. In the case of electron-optical phonon interactions, peaks resulting from phonon absorption satisfying $\ell \hbar \Omega =\Delta _{\text{N},\text{n}}^{{{\text{N}}^{\prime }},{{\text{n}}^{\prime }}}-\hbar \omega _{\mathbf{q}}^{\text{op}}$ are always located to the left of peaks resulting from phonon emission satisfying $\ell \hbar \Omega =\Delta _{\text{N},\text{n}}^{{{\text{N}}^{\prime }},{{\text{n}}^{\prime }}}+\hbar \omega _{\mathbf{q}}^{\text{op}}$. The respective positions of the 1PA, 2PA and 3PA absorption peaks (labeled numerically in Fig. \ref{fig3}(a)) are given in Table 1. From the results in Tab. \ref{tab1}, an interesting observation made by one of the anonymous reviewers is that the distance between the phonon emission peak and the phonon absorption peak of the same $\ell$-photon absorption process is twice the optical phonon energy, i.e. $\ell \left( \hbar {{\Omega }^{\varsigma =1,\text{op}}}-\hbar {{\Omega }^{\varsigma =-1,\text{op}}} \right)=2\hbar \omega _{\mathbf{q}}^{\text{op}}$. This is also easily seen from the relations determining the positions of the phonon emission and absorption peaks mentioned above. Furthermore, peaks resulting from phonon emission exceed those from phonon absorption, indicating that the phonon emission mechanism is greater in strength. The results in Fig. \ref{fig3}a show that the height of the phonon absorption peak has a value equal to about 77\% of the phonon emission peak corresponding to the 1PA and 2PA processes, respectively. 
In Fig. \ref{fig3}(a), another noteworthy point here is the disappearance of the 3PA peak induced by phonon emission (at $\hbar \Omega _{\ell =3}^{\varsigma =1}\approx 38.2\text{meV}$) as the temperature increases. We can explain this disappearance as due to the broadening of the spectral lines of the neighboring peaks upon thermal excitation due to increased temperature. Indeed, we find the positions of the phonon emission/absorption peaks of 1PA and 3PA processes to be $\hbar \Omega _{\ell =3}^{\varsigma =1}\approx 38.2\text{meV}$ and $\hbar \Omega _{\ell =1}^{\varsigma =-1}\approx 41.7\text{meV}$, respectively. These peaks are close together and tend to combine into a single peak as the temperature increases. As the temperature increases, the spectral widths of these individual peaks also increase. As a result, they cannot be observed as clearly as the resonance peaks induced by phonon absorption. We will return to discuss the temperature dependence of the absorption spectra width later. 

\begin{table}[!htb]
\centering
\caption{The positions of the peaks correspond to the transitions in Fig. 3(a).}
\label{tab1}
\begin{ruledtabular}
\begin{tabular}{ccclc}
Labels &
  \begin{tabular}[c]{@{}c@{}}Photon absorption\\ process\end{tabular} &
  \multicolumn{2}{c}{\begin{tabular}[c]{@{}c@{}}Absorption$(-)$/emission$(+)$ \\ phonon process\end{tabular}} &
  \begin{tabular}[c]{@{}c@{}}Location of the peaks $\hbar \Omega$ \\ $\ell \hbar \Omega =\Delta _{\text{N},\text{n}}^{{{\text{N}}^{\prime }},{{\text{n}}^{\prime }}}\pm \hbar \omega _{\mathbf{q}}^{\text{op}}$\end{tabular} \\
(1) & $\ell=3$ & \multicolumn{2}{c}{$-$} & 13.87 meV \\
(2) & $\ell=2$ & \multicolumn{2}{c}{$-$} & 20.82 meV  \\
(3) & $\ell=3$ & \multicolumn{2}{c}{$+$} & 38.25 meV  \\
(4) & $\ell=1$ & \multicolumn{2}{c}{$-$} & 41.61 meV \\
(5) & $\ell=2$ & \multicolumn{2}{c}{$+$} & 56.94 meV \\
(6) & $\ell=1$ & \multicolumn{2}{c}{$+$} & 113.83 meV
\end{tabular}
\end{ruledtabular}
\end{table}

In contrast to the optical phonon interaction, there is no distinction between acoustic phonon absorption and emission processes in the absorption spectra in Fig. \ref{fig3}(b). This has been explained in the previous section by the reason that, assuming the electron-acoustic phonon interactions are elastic, the acoustic phonon energies are considered small and neglected in the argument of the Dirac delta functions \cite{van1,van2}. The three resonance peaks correspond to the 1PA, 2PA, and 3PA processes of LPEMW, satisfying the MPRC condition of the form $\ell \hbar \Omega =\hbar {{\omega }_{B}}+\left( 1-\frac{1}{\lambda } \right)\hbar {{\omega }_{z}}$. This also explains why the peaks of the 2PA and 3PA processes are always located at lower photon energies than the peak of the 1PA process. In addition, the heights of the 2PA and 3PA peaks are approximately 32\% and 6\% of the height of the 1PA peak, respectively. Therefore, it can be seen that the result is that the 1PA process contributes mainly to the total optical absorption process. Additionally, the results of the analysis in Fig. \ref{fig3}b demonstrate that the absorption processes of more than three photons make an even smaller contribution. Consequently, it is advisable to disregard these contributions in the theoretical calculations. 

\begin{figure}[!htb]
    \centering
    \includegraphics[width=1.0\linewidth]{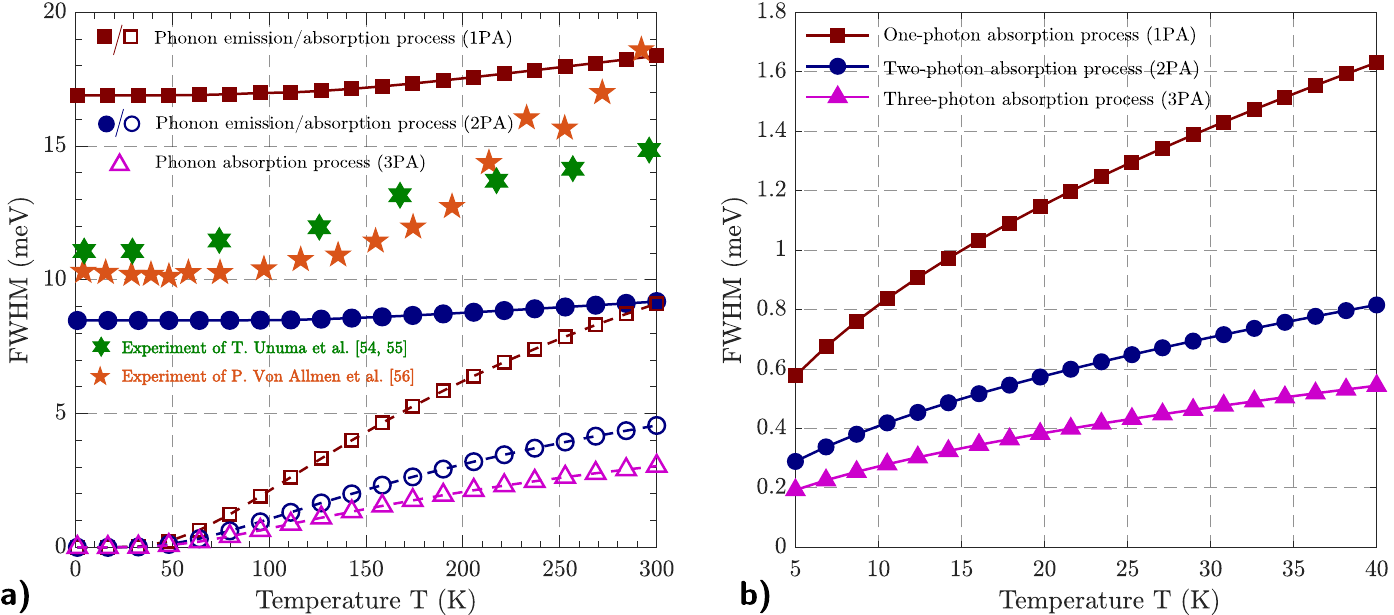}
    \caption{Variation of FWHM as a function of the quantum system temperature for electron-optical phonon interaction (a), and electron-acoustic phonon interaction (b). Here, $L=10$nm, $B=10$T.}
    \label{fig4}
\end{figure}

Now, we will discuss the increase in both the height and width of the spectral lines as temperature increases. The absorption linewidth, also known as Full Width at Half Maximum (FWHM), is estimated using the Profile numerical analysis method as in previous studies on different geometries of confinement potentials in QWs \cite{ptqw, mptqw, hqw, pqw} and graphene \cite{graphene}. There are two reasons for this increase in spectral height, which happens for both optical and acoustic phonon interactions. First of all, from Eqs. \eqref{p}, and \eqref{mep}, we can understand this increase as being caused by the equilibrium distribution functions of electrons and phonons. Indeed, the number of phonons increases exponentially with increasing thermal energy $k_{B}T$, which leads to an increase in the intensity of the absorption peak. For instance, at the observation temperature $T = 77$K, the value of the phonon distribution functions corresponding to the phonon absorption process is $\mathcal{N}_{\mathbf{q}}^{\text{op}}\approx 0.0043$, and the phonon emission process is $\mathcal{N}_{\mathbf{q}}^{\text{op}}+1\approx 1.0043$. Since $\mathcal{N}_{\mathbf{q}}^{\text{op}}$ is much smaller than 1, it shows that the height of the phonon absorption peaks is smaller than that of the phonon emission peaks. Secondly, the phonon absorption or emission process is also represented by Dirac delta functions in Eqs. \eqref{hop}, and \eqref{hac}. These Dirac delta functions are transformed into Lorentzian functions with the spectral broadening calculated by the electron-phonon interaction matrix element. As the temperature of the system increases, the thermal excitation is enhanced, causing the probability of electron transitions in interaction with phonons to increase. This not only increases the intensity or height of the resonance peaks but also increases the width of the absorption spectral line with temperature. We present the results of the temperature dependence of the spectral line width or FWHM in Fig. \ref{fig4}. 

The effect of temperature on the FWHM for the both of two interaction mechanisms in Morse QWs is shown in Fig. \ref{fig4}. As discussed above, as temperature rises, the quantity of phonons increases, so elevating the likelihood of electron-phonon scattering probability, which also results in the widening of absorption peaks and enhances phonon-assisted transitions. The FWHM increases with temperature with different rules depending on the mechanism of optical phonon absorption or emission. In the case of optical phonon emission, the FWHM (W) depends on temperature in an exponential law for both
one-, and two- photon absorption, with its magnitude proportional to the phonon distribution function as follows 
\begin{align}\label{wt1}
    \text{W}_{\varsigma =1}^{\left( \text{op} \right)}\left( T \right)=\alpha _{\varsigma =1}^{\left( \text{op} \right)}+\beta _{\varsigma =1}^{\left( \text{op} \right)}\mathcal{N}_{\mathbf{q}}^{\text{op}}\left( T \right), 
\end{align}
here, $\beta _{\varsigma =1}^{\left( \text{op} \right)}$ is the optical-phonon broadening parameter, while $\alpha _{\varsigma =1}^{\left( \text{op} \right)}$ is the stable parameter of the FWHM, which is primarily attributed to impurities and acoustic phonon interactions at extremely low temperatures \cite{ptqw}. Meanwhile, for optical phonon absorption, the FWHM depends on temperature with a rule such that its magnitude is proportional to the square root of the optical phonon distribution function as follows 
\begin{align}\label{wt2}
    \text{W}_{\varsigma =-1}^{\left( \text{op} \right)}\left( T \right)=\gamma _{\varsigma =-1}^{\left( \text{op} \right)}\sqrt{\mathcal{N}_{\mathbf{q}}^{\text{op}}\left( T \right)}.
\end{align}
In the case of acoustic phonon interactions, due to the approximation of the phonon distribution function \cite{van3}, i.e. $\mathcal{N}_{\mathbf{q}}^{\text{ac}}+1\approx \mathcal{N}_{\mathbf{q}}^{\text{ac}}\approx {{{k}_{B}}T}/{\hbar \omega _{\mathbf{q}}^{\text{ac}}}\;$, the temperature dependence of the FWHM has the form 
\begin{align}\label{wact}
    {{\text{W}}^{\left( \text{ac} \right)}}\left( T \right)={{\alpha }^{\left( \text{ac} \right)}}+{{\beta }^{\left( \text{ac} \right)}}\sqrt{T},
\end{align}
here, the fitted parameters, $\alpha _{\varsigma =1}^{\left( \text{op} \right)}$, $\beta _{\varsigma =1}^{\left( \text{op} \right)}$, $\gamma _{\varsigma =-1}^{\left( \text{op} \right)}$, ${{\alpha }^{\left( \text{ac} \right)}}$ and ${{\beta }^{\left( \text{ac} \right)}}$ are given in Tab. \ref{tab2}. The rules expressed by Eqs. \eqref{wt1}, \eqref{wt2} and \eqref{wact} are interpolated as indicated by the thermal broadening mechanism presented in the previous experimental measurements of D. Gammon \textit{et al} \cite{gammon} and D.A.B. Miller \textit{et al} \cite{gionggammon}, and theoretical studies in other QWs \cite{gqw, hqw, tqw, ptqw, mptqw, pqw1}. In particular, in Fig. \ref{fig4}a, for the electron-optical phonon interactions, our calculated results give good agreement with the experimental observation data of T. Unuma \textit{et al.} \cite{unuma1, unuma2} in a single square QW and P. Von Allmen \textit{et al.} \cite{vonexp} in multi-QW structures. However, our FWHM calculations for the 1PA process are still slightly larger than these experimental data. This can be explained by the fact that the influence of the magnetic field and the electron confinement effect in the Morse QW are stronger than in the square QW. 

\begin{table}[!htb]
\centering
\begin{ruledtabular}
\begin{tabular}{cccccc}
\textbf{Symbol} &
  $\alpha _{\varsigma =1}^{\left( \text{op} \right)}$ &
  \multicolumn{1}{c}{$\beta _{\varsigma =1}^{\left( \text{op} \right)}$} &
  \multicolumn{1}{c}{$\gamma _{\varsigma =-1}^{\left( \text{op} \right)}$} &
  \multicolumn{1}{c}{${{\alpha }^{\left( \text{ac} \right)}}$} &
  \multicolumn{1}{c}{${{\beta }^{\left( \text{ac} \right)}}$} \\ \hline
\textbf{Unit} & \multicolumn{4}{c}{meV}           & $\text{meV}\cdot \text{K}^{-1/2}$ \\ 
\midrule
\textbf{1PA}  & 16.9     & 4.53     & 15.9 & 0.64 & 1.76                              \\
\textbf{2PA}  & 8.46     & 2.21     & 7.9  & 0.32 & 0.88                              \\
\textbf{3PA}  & $\cdots$ & $\cdots$ & 5.3  & 0.32 & 0.58     \\     
\end{tabular}
\end{ruledtabular}
\caption{The values of the fitted parameters, $\alpha _{\varsigma =1}^{\left( \text{op} \right)}$, $\beta _{\varsigma =1}^{\left( \text{op} \right)}$, $\gamma _{\varsigma =-1}^{\left( \text{op} \right)}$, ${{\alpha }^{\left( \text{ac} \right)}}$ and ${{\beta }^{\left( \text{ac} \right)}}$ for different phonon interaction mechanisms correspond to each $\ell$-photon absorption process from Fig. \ref{fig4}.}
\label{tab2}
\end{table}

\begin{figure}[!htb]
    \centering
    \includegraphics[width=1\linewidth]{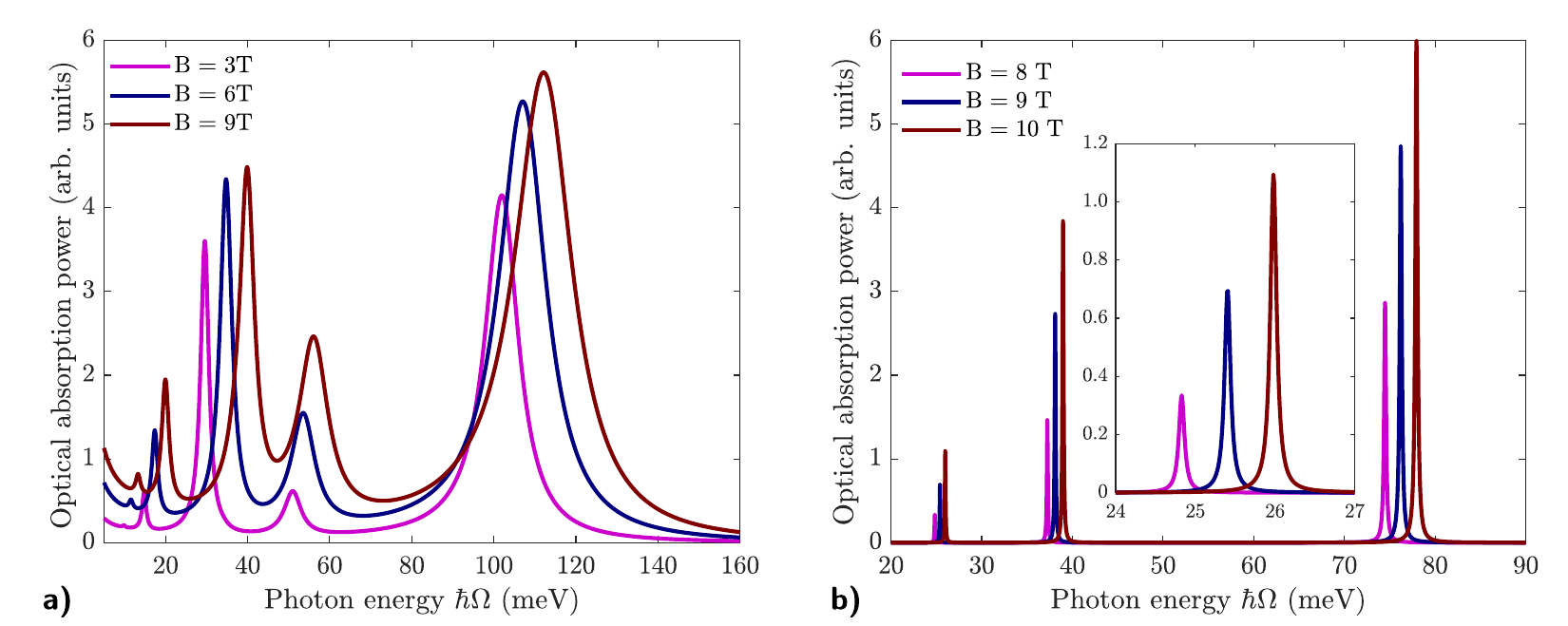}
    \caption{Calculated incident photon energy dependence of the OAP of Morse QWs on several different values of the external magnetic field in both of the optical phonon (a) and acoustic phonon (b) interactions. Here, $L=10$nm, $T=150K$ for optical phonon interactions, and  $T=5K$ for acoustic phonon interactions.}
    \label{fig5}
\end{figure}
\begin{figure}[!htb]
    \centering
    \includegraphics[width=1\linewidth]{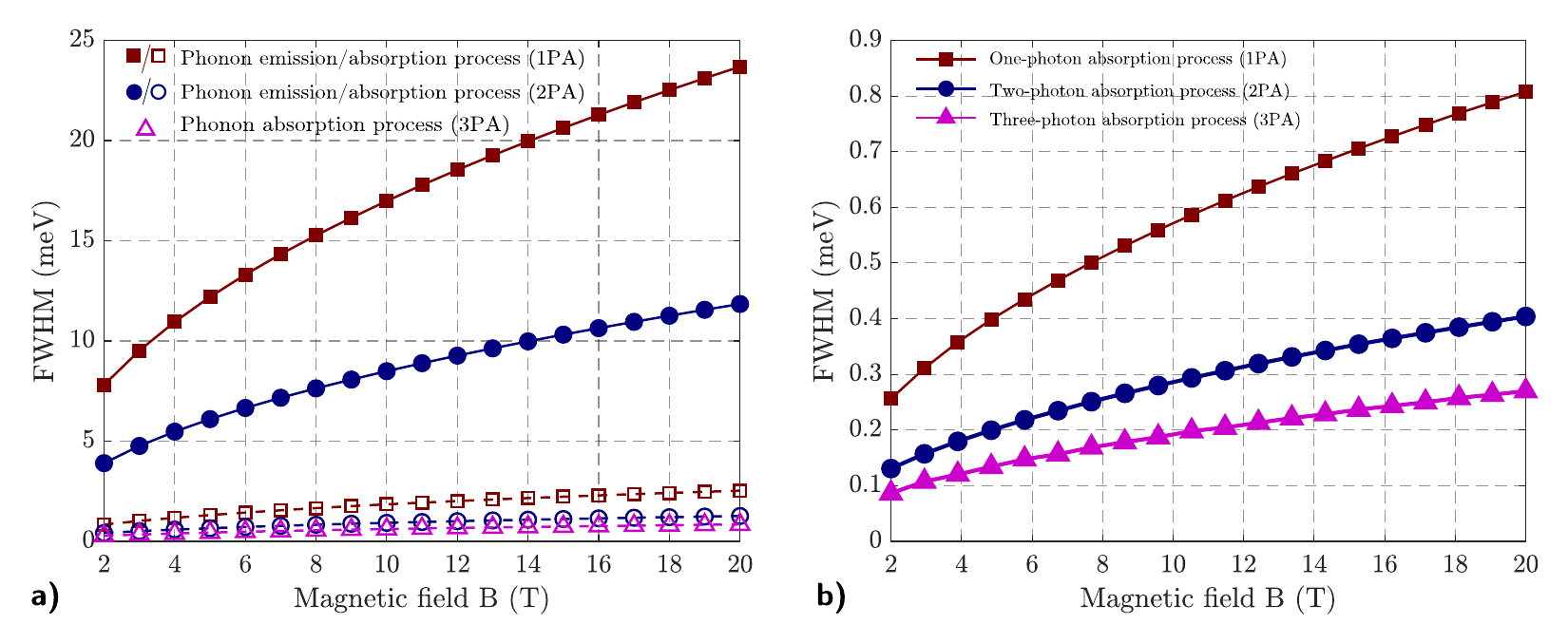}
    \caption{The magnetic field influences the FWHM in the Morse QWs at (a) $T = 150$K for optical phonon interactions and (b) $T = 5$K for acoustic phonon interactions, respectively.}
    \label{fig6}
\end{figure}

Magnetic field is considered as one of the significant external influences on the magneto-optical properties of materials, specifically the optical absorption spectra of electrons. In Fig. \ref{fig5}, the OAP in the Morse QW is shown as a function of the photon energy of LPEMW for several values of the external magnetic field in both of the two interaction mechanisms. Indeed, the energy spectra of electrons in two-dimensional QWs under the influence of a magnetic field exhibits Landau splitting. As the magnetic field increases, the cyclotron frequency also increases, thus increasing the energy separation $\Delta _{\text{N},\text{n}}^{{{\text{N}}^{\prime }},{{\text{n}}^{\prime }}}$. Based on the analysis related to MPRC $\ell \hbar \Omega =\Delta _{\text{N},\text{n}}^{{{\text{N}}^{\prime }},{{\text{n}}^{\prime }}}+\varsigma \hbar \omega _{\mathbf{q}}^{\nu }$ (recall, $\varsigma =1$ for phonon emission, $\varsigma =-1$ for phonon absorption, and $\varsigma =0$ for acoustic phonon interactions), we can observe the consequences of this through a blue-shift behavior of the absorption spectra in Fig. \ref{fig5}, namely the resonance peaks will shift to the right towards the higher-energy region. In addition, we also found that as the magnetic field increases, the value of the absorption peak also increases. These results are also consistent with the conclusions shown in previous works \cite{pqw,pqw1,pqw2,hqw, ptqw, mptqw}, where they only considered the interaction between electrons and optical phonons. This growth trend can be explained as follows: (i) quantitatively, the OAP is proportional to $\mathscr{P}_0$, and since $\mathscr{P}_0$ is inversely proportional to $\ell_B^6$, the OAP becomes proportional to $B^6$, and (ii) physically, the magnetic field induces parabolic confinement in the directions perpendicular to its axis \cite{btaorathegiam}. Thus, when $\bf{B}$ is aligned with the Oz axis, the magnetic field confines the x-direction, enhancing the electron confinement effect in the Morse QW, which leads to stronger OAP. The blue-shift and intensity enhancement behavior of the resonance peaks with magnetic field is also observed in nonlinear optical rectification, total absorption and second harmonics generation coefficients in different QWs by the compact density matrix method \cite{ugan2,ugan4,gqws}. In addition, under the influence of broadening Landau level mechanism \cite{shon1998quantum}, the magnetic field also enhances the Lorentzian width, thereby broadening the absorption spectral line, or FWHM. We present the influence of magnetic field on FWHM as shown in Fig. \ref{fig6}.

\begin{table}[!htb]
\centering
\begin{ruledtabular}
\begin{tabular}{ccccl}
\multirow{3}{*}{\begin{tabular}[c]{@{}c@{}}$\mu$\\ ($\text{meV}\cdot \text{T}^{-1/2}$)\end{tabular}} &
  \multicolumn{4}{c}{$\text{FWHM} = \mu \times \sqrt{B}$ (meV; T)} \\ \cline{2-5} 
 &
  \multicolumn{2}{c}{\textbf{\begin{tabular}[c]{@{}c@{}}Optical phonon \\ interactions\end{tabular}}} &
  \multicolumn{2}{c}{\textbf{\begin{tabular}[c]{@{}c@{}}Acoustic phonon \\ interactions\end{tabular}}} \\ \cline{2-5} 
 &
  \begin{tabular}[c]{@{}c@{}}$\varsigma = 1$\end{tabular} &
  \begin{tabular}[c]{@{}c@{}}$\varsigma = -1$\end{tabular} &
  \multicolumn{2}{c}{$\varsigma = 0$} \\ 
  \midrule
\textbf{1PA} & 5.22 & 0.52 & \multicolumn{2}{c}{0.18} \\
\textbf{2PA} & 2.61 & 0.26 & \multicolumn{2}{c}{0.12} \\
\textbf{3PA} & 1.74 & 0.17 & \multicolumn{2}{c}{0.06} \\
\end{tabular}
\end{ruledtabular}
\caption{The values of the fitted parameter $\mu$ for different phonon interaction mechanisms correspond to each $\ell$-photon absorption process.}
\label{tab3}
\end{table}

Fig. \ref{fig6} describes magnetic field dependence of the FWHM for different mechanisms of phonon interaction. From the figure, it can be seen that the FWHM increases monotonically with the square root of the magnetic field $\text{FWHM}=\mu \times \sqrt{B}$ (meV; T) in both interaction mechanisms but with different values of the factor of proportionality $\mu$. Qualitatively, this law is similar to that shown in previous studies \cite{van1, van2} using linear response theory in two-dimensional electronic systems. Quantitatively, the $\mu$ coefficients characterizing the magnitude of the FWHM for each phonon and photon absorption/emission process are given in Tab. \ref{tab3}. Here, the magnetic field-dependent expression of the FWHM in Morse QWs does not have a free term as in the hyperbolic QW \cite{hqw,oapwt1} and special symmetric QW \cite{oapwt2}. However, this expression is consistent with the results obtained in the Pöschl-Teller \cite{ptqw}  and the modified Pöschl-Teller \cite{mptqw} QWs with a slightly smaller $\mu$ factor corresponding to the one- and two-photon absorption processes in the optical phonon interaction. Our calculations also show that the FWHM of the 3PA process is about 32\% of the FWHM of the 1PA process. This demonstrates that the nonlinear absorption processes have an important contribution that cannot be ignored when studying the electron transport phenomena in low-dimensional semiconductor systems. In addition, the calculations in Fig. \ref{fig6}(b) show that the FWHM in the case of acoustic phonon interactions (which are considered dominant at low temperatures) has relatively small values. Our estimate also strongly supports similar results obtained on the magnetic field and temperature dependence of the cyclotron-resonance linewidth due to the electron-acoustic phonon interaction in multi-quantum-well structures using the deformation-potential approximation performed by M. Singh \cite{pna}. 
Indeed, our (in Morse QWs) and M. Singh's estimates (in multi-quantum-well structures) for the FWHM in the electron-acoustic phonon interaction are quite small in the extremely low temperature region (less than 10K), only about 0.1 - 1.6 meV. This may be one of the main reasons why it is difficult to detect the magneto-phonon resonance spectral lines in the acoustic phonon interaction mechanism. Therefore, to our present knowledge, there are no experimental observations to verify the validity of the above theoretical result. We hope that this theoretical suggestion will provide further rational evidence and useful criteria for future advances in theory and experiment. An interesting suggestion from one of the anonymous reviewers is to consider the calculation of the FWHM when the magnetic field is zero. However, this work only focuses on the magneto-phonon resonance in Morse QWs when the electron-phonon system is placed in an external magnetic field and a strong electromagnetic wave. In the absence of a magnetic field, the FWHM of the electron-phonon resonance can also be calculated numerically using the Profile method in a similar way \cite{graphene} and gives quite small values as shown in our very recent study in a quantum well model with asymmetric semi-parabolic potentials \cite{pqw}.

\begin{figure}[!htb]
    \centering
    \includegraphics[width=1\linewidth]{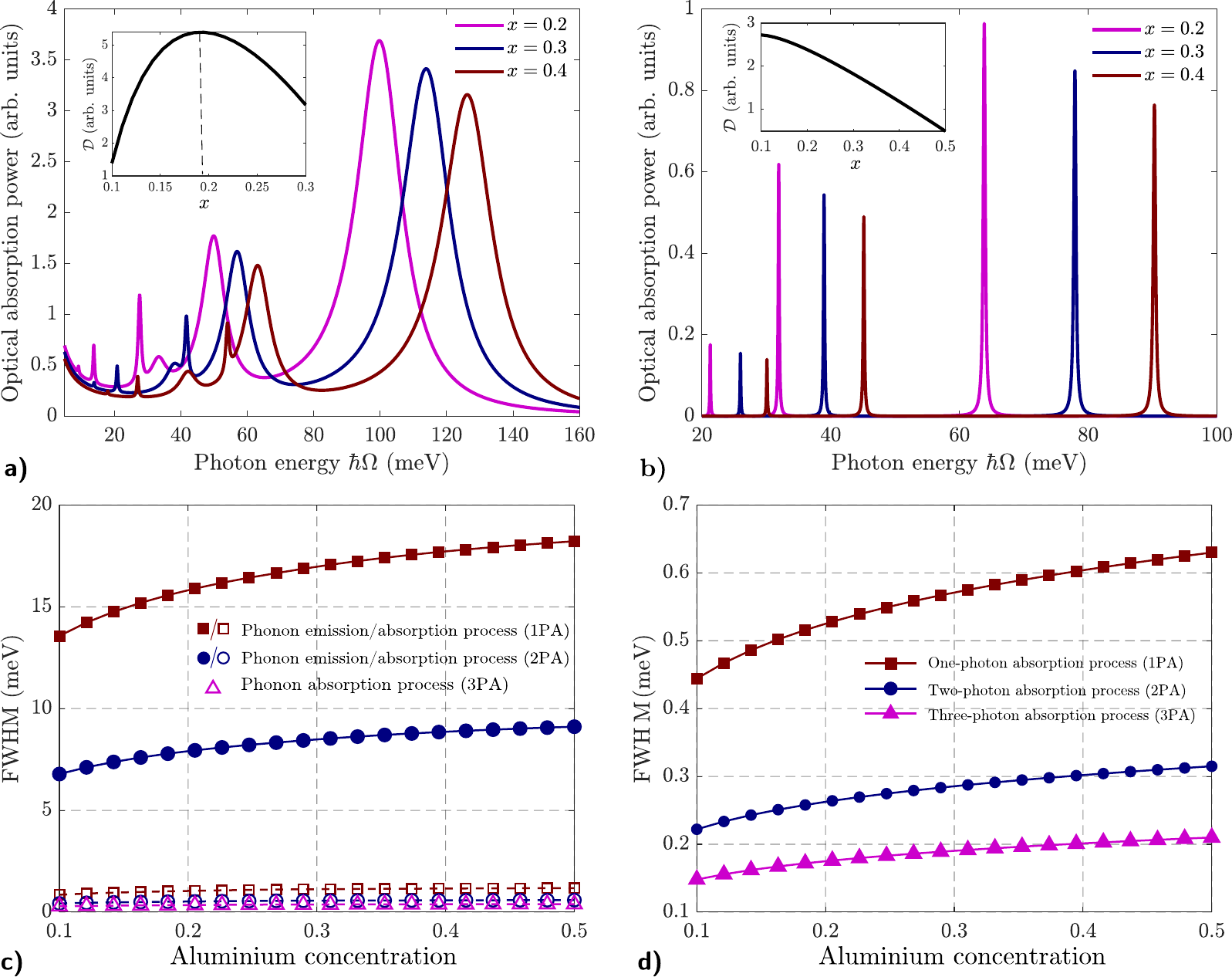}
    \caption{Optical absorption spectra of electrons [(a), (b)] and FWHM [(c), (d)] with varying aluminum doping concentration corresponding to optical phonon [(a), (c)] and acoustic phonon [(b), (d)] interaction mechanisms. Here, $L = 10$nm, $B = 10$T, and T = 150K in Figs. (a), (c), T = 5K in Figs. (b), (d). In subfigures in Figs. \ref{fig7}(a), and \ref{fig7}(b), the black curve shows the dependence of the ${{\mathcal{D}}^{\left( \text{op/ac} \right)}}$-factor on the aluminum concentration.}
    \label{fig7}
\end{figure}

In addition to the evaluation of the influence of external fields (temperature, magnetic field) on the optical properties, we are also interested in the influence of material parameters on the absorption spectra as well as the FWHM in Morse QWs. Two important parameters that can influence the profile of Morse QWs can be seen in Fig. \ref{fig1}, including the well width $L$ and the aluminum concentration $x$ (related to the well depth). Firstly, the role of aluminum concentration is related to the height of the barrier (recall that ${{U}_{0}}=0.6\times \left( 1155\times x+370\times {{x}^{2}} \right)\left( \text{meV} \right)$). In Figs. \ref{fig7}a and \ref{fig7}b, it can be seen that, as the aluminum concentration $x$ increases, the height of the resonance peaks in the absorption spectra decreases as the potential well becomes deeper. Quantitatively, from the theoretical results in Eqs. \eqref{apop} and \eqref{apac}, we derive the $\mathcal{D}$-factor of the form ${{\mathcal{D}}^{\left( \text{op/ac} \right)}}=\mathscr{P}_{0}^{\left( \text{op/ac} \right)}{{\Lambda }_{\text{N},\text{n}}}{{\mathcal{G}}_{{{\text{n}}^{\prime }},{{\text{n}}^{\prime {\prime }}}}}{{\mathcal{Q}}_{{{\text{N}}^{\prime {\prime } }},\text{N}}}$ as a quantity characterizing the intensity of the absorption peaks. The graphs showing the dependence of the $\mathcal{D}$-factor on the aluminum concentration are given as subfigures in Figs. \ref{fig7}(a) and \ref{fig7}(b). For the  electron-optical phonon interaction in Fig. \ref{fig7}(a), as the aluminum concentration $x$ increases, the $\mathcal{D}$-factor first rises, reaches its peak at $x \approx 0.19$, and thereafter diminishes with further increases in $x$. Meanwhile, for the electron-acoustic phonon interaction in Fig. \ref{fig7}(b), the value of the $\mathcal{D}$-factor decreases monotonically as the aluminum concentration increases. This indicates that as the well depth increases, the peak intensity diminishes, consistent with prior findings in a finite square-shaped \cite{pqw2} and semi-parabolic \cite{smpqw} QWs. The blue-shifted behavior of the absorption spectra is also observed here as the aluminum concentration increases. The cause of this occurrence is because the aluminum concentration has a strong influence on the height of the potential well $U_0$, which in turn indirectly affects the factor $\lambda$ and the confinement frequency $\omega_z$ (recall, $\lambda \sim \sqrt{{{U}_{0}}}$, and ${{\omega }_{z}}\sim \sqrt{{{U}_{0}}}$). This leads to an increase in the energy separation since  $\Delta _{\text{N},\text{n}}^{{{\text{N}}^{\prime }},{{\text{n}}^{\prime }}}\sim \left( 1-{{\lambda }^{-1}} \right)\hbar {{\omega }_{z}}$. From the MPRC, we see that the position of the resonance peaks is a function of the energy separation $\Delta _{\text{N},\text{n}}^{{{\text{N}}^{\prime }},{{\text{n}}^{\prime }}}$, so the resonance peaks will shift towards higher photon energies. The well depth increases significantly with increasing aluminum concentration, which also enhances the influence of the quantum confinement effect of Morse QWs on the FWHM. 
In Figs. \ref{fig7}(c) and \ref{fig7}(d), it can be seen that the FWHM increases monotonically nonlinearly as the aluminum concentration increases from 0.1 to 0.5, whereas this rule is linear when previously investigated in a finite square-shaped \cite{pqw2} and semi-parabolic \cite{smpqw} QWs. The nonlinear dependence of the FWHM on aluminum concentration found here is consistent with the result for the semi-parabolic plus semi-inverse squared QW \cite{afqw} but is slightly larger in the case of optical phonon interactions. This demonstrates that the quantum confinement and electron-phonon interactions in the Morse QW are stronger than in other confinement potential profiles. 

\begin{figure}[!htb]
    \centering
    \includegraphics[width=1\linewidth]{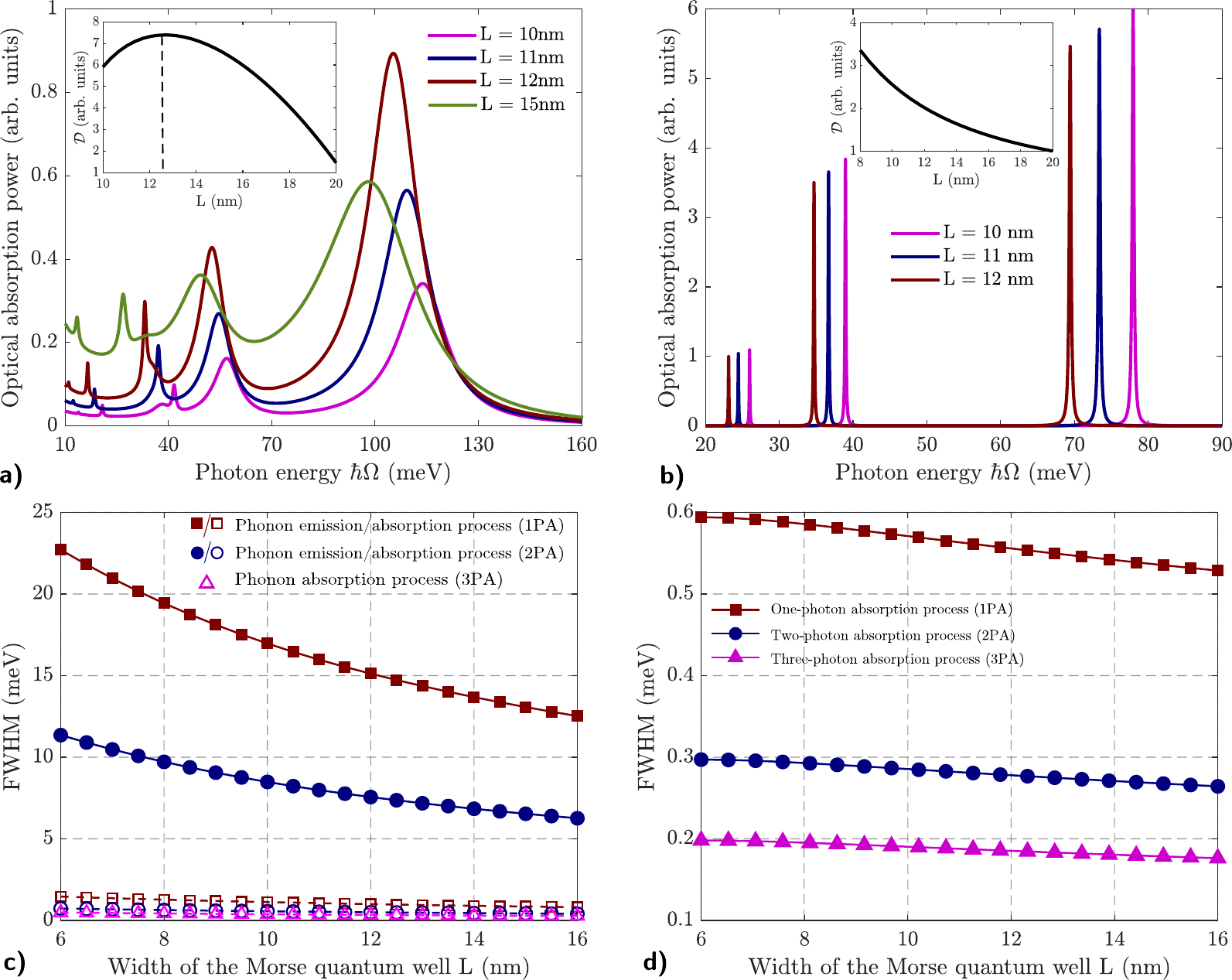}
    \caption{Optical absorption spectra of electrons [(a), (b)] as a function of the incident photon energy of LPEMW and the dependence of FWHM [(c), (d)] on the width of Morse QWs corresponding to optical phonon [(a), (c)] and acoustic phonon [(b), (d)] interaction mechanisms. Here, $x = 0.3$, $B = 10$T, and T = 150K in Figs. (a), (c), T = 5K in Figs. (b), (d). In subfigures in Figs. \ref{fig8}a, and \ref{fig8}b, the black curve shows the dependence of the ${{\mathcal{D}}^{\left( \text{op/ac} \right)}}$-factor on the well-width of Morse QWs.}
    \label{fig8}
\end{figure}

Along with the aluminum concentration, the width $L$ of the confinement potential is also a vital extrinsic parameter that strongly influences optical absorption in Morse QWs. Fig. \ref{fig8} presents our estimates of the optical absorption spectra and FWHM with respect to the variation of the confinement width $L$ of the Morse QWs in both phonon interaction mechanisms. In contrast to the results obtained in Figs. \ref{fig5}, \ref{fig7}a, and \ref{fig7}b, as the width $L$ of the Morse QW increases, the absorption spectra in Figs. \ref{fig8}a and \ref{fig8}b in both interaction mechanisms show a shift of the resonance peaks to the left towards lower photon energies (red-shift behavior). The reason behind the red-shift behavior of the resonance peak is due to the decrease in energy separation as the width of the potential well increases. Indeed, mathematically, since the confinement frequency is inversely proportional to $L$, while the factor $\lambda$ is directly proportional to $L$, the energy separation for the extreme quantum limit ($\text{N}=\text{n}=0;{{\text{N}}^{\prime }}={{\text{n}}^{\prime }}=1$) \cite{van3} depend on $L$ as $\Delta _{0,0}^{1,1}\sim {\left( aL-b \right)}/{{{L}^{2}}}\;$, with $a={{\hbar }^{2}}\sqrt{{2{{U}_{0}}}/{{{m}_{e}}}\;}$, and $b={{{\hbar }^{3}}}/{{{m}_{e}}}\;$. Hence the energy separation tends to increase as the width increases. Furthermore, as the width of the potential well increases, the QW widens and approaches the characteristics of the bulk material. This weakens the influence of the quantum confinement effect and is the main cause of the decrease in the resonance peak intensity of the OAP (see Figs. \ref{fig8}(a), \ref{fig8}(b)) and FWHM (see Figs. \ref{fig8}(c), \ref{fig8}(d)). To investigate the dependence of the absorption peak intensity on the width $L$ of the Morse QW, we present the sub-figures in Figs. \ref{fig8}(a), and \ref{fig8}(b), with the black curve showing the dependence of the $\mathcal{D}$-factor characterizing the peak intensity on the width $L$. As can be seen in \ref{fig8}(a) (optical phonon interactions), the value of $\mathcal{D}$-factor increases to a maximum at $L\approx 12.4\text{nm}$, then decreases rapidly as $L$ continues to increase. On the other hand, for the acoustic phonon interaction, as $L$ increases, the $\mathcal{D}$-factor decreases rapidly, leading to a decrease in the intensity of the absorption peaks as shown in Fig. \ref{fig8}(b). This is also consistent with the findings in the finite semi-parabolic QW \cite{smpqw} with optical phonon interaction, but there the authors found the $\mathcal{D}$-factor to have a maximum at $L\approx 17.4\text{nm}$, which is larger than the value found ($L\approx 12.4\text{nm}$) in this study. 
To provide a quantitative prediction for the reduced behavior of the FWHM under the broadening trend of the Morse QW for the optical phonon interaction mechanism, we propose an expression of the form ${{\text{W}}^{\left( \text{op} \right)}}\left( L \right)={{{\varpi }^{\left( \text{op} \right)}}}/{\sqrt{L}}\;$. Where, ${{\varpi }^{\left( \text{op} \right)}}= 53.55, 26.77,$  $\text{meV}\cdot \text{n}{{\text{m}}^{{1}/{2}\;}}$  for transitions requiring the electron to emit one phonon, while ${{\varpi }^{\left( \text{op} \right)}}= 3.55, 1.76, 1.17$ $\text{meV}\cdot \text{n}{{\text{m}}^{{1}/{2}\;}}$ for transitions requiring the electron to absorb one phonon for the 1PA, 2PA, 3PA processes, respectively. The $L$-dependence of the FWHM and the parameters found are in good agreement with the results of H. V. Phuc \textit{et al.} \cite{pqw3} in finite parabolic QWs, but with a slightly larger parameter ${{\varpi }^{\left( \text{op} \right)}}$. This demonstrates that the influence of the quantum confinement effect on the optical properties in MQWs is larger than in parabolic QWs. This rule differs from the prediction of L. V. Tung \textit{et al.} \cite{smpqw} in finite semi-parabolic QWs where FWHM is inversely proportional to ${{L}^{{3}/{2}\;}}$. For the acoustic phonon interaction mechanism, our predicted expression takes the form of a linear decay function ${{\text{W}}^{\left( \text{ac} \right)}}\left( L \right)=\wp -{{\varpi }^{\left( \text{ac} \right)}}\times L$, with ${{\varpi }^{\left( \text{ac} \right)}} = 7, 3.5, 2.3$ $10^{-3} \text{meV}\cdot \text{n}{{\text{m}}^{-1}}$, and $\wp = 0.64, 0.32, 0.21$ meV for the 1PA, 2PA, 3PA processes, respectively. Fig. \ref{fig8}d illustrates a minor decrease in FWHM with respect to $L$, with the ${{\varpi }^{\left( \text{ac} \right)}}$-parameter being rather little. This indicates that $L$ has a restricted impact on the interactions between electrons and acoustic phonons in Morse QWs. 

\section{Conclusions}
In this work, the effects  magnetic field, temperature and material structure parameters on OAP and FWHM of Morse QWs generated by $\text{GaAs}/\text{A}{{\text{l}}_{x}}\text{G}{{\text{a}}_{1-x}}\text{As}$ heterostructure under the influence of LPEMW have been examined in detail. The results are considered in both optical and acoustic phonon interaction mechanisms and take into account the MPA of LPEMW. The 1PA peaks are always larger and occur to the right of the 2PA peaks, which in turn are larger and appear to the right of the 3PA peaks. The intensity of the 1PA peaks (linear absorption) is the highest, while the nonlinear absorption peaks (2PA and 3PA) have lower values, with the intensity of the 3PA peaks being smaller than that of the 2PA peaks. In the optical phonon interaction mechanism, peaks from phonon absorption are always smaller and occur to the left of those from phonon emission. The positions of the resonance peaks follow the magneto-phonon resonance condition and are independent of temperature. However, when the magnetic field is applied along the growth direction and the aluminum concentration increases, the absorption spectra exhibits a blue shift. Conversely, a red shift in the absorption spectra occurs as the QW width increases. Changes in Morse parameters (aluminum concentration, well width), and magnetic field) cause these shift behaviors due to variations in separation energy. This result is highly valuable for applications involving the use of LPEMW to detect magneto-phonon resonance processes and determine electron energy subbands in QWs by measuring the spacing between resonance peaks. Although thermal excitations do not affect the positions of the resonance peaks, they tend to increase their intensity. In addition to thermal excitation, the intensity of the absorption peaks also increases under the influence of an external magnetic field. Conversely, when the width and depth (related to the aluminum concentration) of the QW increase beyond a critical value, the intensity of the absorption peak decreases. 

By employing the numerical method, we obtained numerical values for the FWHM of the resonance peaks in the optical absorption spectra. Our results indicate that the FWHM increases with rising magnetic field strength, temperature, and aluminum concentration, but decreases as the QW width expands. Although the nonlinear absorption processes (2PA, 3PA) exhibit smaller FWHM values compared to the linear absorption process (1PA), they still demonstrate significant strength, highlighting their importance in studying the optical absorption spectra of electrons. Moreover, the observed thermal broadening of the FWHM is consistent with previous findings in other QW models and with experimental observations from square QWs and multi-QWs. It is important to note that the FWHM of the resonance peaks associated with acoustic phonon interactions is significantly smaller than that of optical phonon interactions. Nevertheless, we present expressions predicting the dependence of the FWHM on magnetic field strength, aluminum concentration, and QW width, addressing gaps in previous theoretical studies. However, there are currently no direct experimental results available for comparison with our theoretical FWHM calculations in Morse QWs. We hope that future experimental work will soon validate our predictions. Overall, our theoretical study of the OAP and FWHM of the resonance peaks demonstrates that the Morse QW model offers promising magneto-optical properties, making it a strong candidate for future applications in optoelectronic devices. 

\begin{acknowledgments}
We thank the anonymous referees for their valuable comments and suggestions. One of the authors (A-T.T) wishes to express gratitude to Mr. Quang-Huy Nguyen Dang (VNU Vietnam Japan University) and Dr. Nguyen Thuy Trang (VNU University of Science) for their helpful discussions during the course of this work. We sincerely thank Dr. Nguyen Viet Hung (Hanoi University of Science and Technology) for his diligent proofreading of this paper.
\end{acknowledgments}
\appendix 
\section{Evaluate the integrals in Eqs. \eqref{hop}, and \eqref{hac}}
In this appendix, we present the values of several integrals related to Laguerre polynomials characteristic of MPA processes in Eqs. \eqref{hop}, and \eqref{hac}. Now, we rewrite the integrals as follows   
\begin{align}
    \mathscr{I}_{{{\text{N}}^{\prime }}\text{,}{{\text{N}}^{\prime \prime }}}^{\left( \ell  \right)}=\int\limits_{0}^{+\infty }{{{u}^{\ell }}{{\left| {{\mathcal{J}}_{{{\text{N}}^{\prime }}\text{,}{{\text{N}}^{\prime \prime }}}}\left( u \right) \right|}^{2}}du}=\frac{{{\text{N}}^{\prime \prime }}!}{{{\text{N}}^{\prime }}!}\int\limits_{0}^{+\infty }{\exp \left( -u \right){{u}^{{{\text{N}}^{\prime }}-{{\text{N}}^{\prime \prime }}+\ell }}{{\left[ \mathcal{L}_{{{\text{N}}^{\prime \prime }}}^{{{\text{N}}^{\prime }}-{{\text{N}}^{\prime \prime }}}\left( u \right) \right]}^{2}}du}.
\end{align}

From Eqs. (A1), and (A4) of Ref. \cite{tp1}, and Eq. (A1) of Ref. \cite{tp2}, we get 
\begin{align}
  & \mathscr{I}_{{{\text{N}}^{\prime }}\text{,}{{\text{N}}^{\prime \prime }}}^{\left( 0 \right)}=1, \\ 
 & \mathscr{I}_{{{\text{N}}^{\prime }}\text{,}{{\text{N}}^{\prime \prime }}}^{\left( 1 \right)}={{\text{N}}^{\prime }}+{{\text{N}}^{\prime \prime }}+1, \\ 
 & \mathscr{I}_{{{\text{N}}^{\prime }}\text{,}{{\text{N}}^{\prime \prime }}}^{\left( 2 \right)}={{\text{N}}^{\prime \prime }}^{2}+4{{\text{N}}^{\prime \prime }}{{\text{N}}^{\prime }}+3{{\text{N}}^{\prime \prime }}+{{\text{N}}^{\prime }}^{2}+3{{\text{N}}^{\prime }}+2, \\ 
 & \mathscr{I}_{{{\text{N}}^{\prime }}\text{,}{{\text{N}}^{\prime \prime }}}^{\left( 3 \right)}={{\text{N}}^{\prime \prime }}^{3}+9{{\text{N}}^{\prime \prime }}^{2}{{\text{N}}^{\prime }}+6{{\text{N}}^{\prime \prime }}^{2}+9{{\text{N}}^{\prime \prime }}{{\text{N}}^{\prime }}^{2}+18{{\text{N}}^{\prime \prime }}{{\text{N}}^{\prime }}+11{{\text{N}}^{\prime \prime }}+{{\text{N}}^{\prime }}^{3}+6{{\text{N}}^{\prime }}^{2}+11{{\text{N}}^{\prime }}+6. 
\end{align}
\bibliography{main}
\end{document}